\documentclass[fleqn,usenatbib]{mnras}

\usepackage[T1]{fontenc}

\DeclareRobustCommand{\VAN}[3]{#2}
\let\VANthebibliography\thebibliography
\def\thebibliography{\DeclareRobustCommand{\VAN}[3]{##3}\VANthebibliography}

\usepackage{graphicx}	
\usepackage{amsmath}	
\usepackage{amssymb}	

\title[Hydrodynamic models of R Hya]{Hydrodynamic models of pulsation period evolution in R Hydrae}

\author[Yu. A. Fadeyev]{
Yuri A. Fadeyev,$^{1}$\thanks{E-mail: fadeyev@inasan.ru}
\\
$^{1}$Institute of Astronomy, Russian Academy of Sciences,  119017, Pyatnitskaya str., 48, Moscow, Russia
}

\date{Accepted XXX. Received YYY; in original form ZZZ}

\pubyear{2015}

\begin{document}
\label{firstpage}
\pagerange{\pageref{firstpage}--\pageref{lastpage}}
\maketitle

\begin{abstract}
Pulsation period evolution during the helium--shell flash in the Mira variable R~Hya is investigated
using consistent stellar evolution and non--linear stellar pulsation computations.
The initial and time--dependent inner boundary conditions for the equations of radiation hydrodynamics
describing non--linear stellar oscillations were determined using a grid of TP--AGB model sequences
with initial masses on the main sequence $\mathbf{1.5M}_{\sun}\le M_\mathrm{ZAMS}\le 5.0M_{\sun}$
and the initial metallicity $Z=0.014$.
The setup of initial conditions for hydrodynamic models corresponds to $\approx 100$~yr prior to
the maximum of the helium--shell luminosity and ensures that the stellar envelope of the evolution
model is under both hydrostatic and thermal equilibrium.
Solution of the equations of hydrodynamics allowed us to determine the temporal variation of
the pulsation period $\Pi(t)$ during $\approx 500$~yr.
Within this time interval R~Hya is a fundamental mode pulsator.
The period temporal dependencies $\Pi(t)$ calculated for the AGB star models at the beginning of
the third dredge--up phase and with masses $4.4M_{\sun}\loa M\loa 4.5M_{\sun}$ are in agreement
with observational estimates of the period of R~Hya obtained during last two centuries.
The mean radius of R~Hya pulsation models at the end of the XX century
($470 R_{\sun}\loa\bar{R}\loa 490 R_{\sun}$) agrees with observational estimates obtained using
the interferometric angular diameter measurements.
\end{abstract}




\section{Introduction}

First observations of R~Hya were conducted by Hevelius in 1662 and Montanary in 1670 \citep{A1869},
whereas discovery of variability is attributed to Maraldi in 1704 \citep{MH1922}.
Regular observations of R~Hya were started in the middle of XIX century and great attention to this
variable star was attracted due to detection of the decreasing period of its light variations
\citep{G1882}.
Period decrease in R~Hya was reviewed by \citet{Ch1896} and \citet{P1936} but the most recent and
comprehensive analysis of period changes can be found in the paper by \citet{ZBM2002}.
In particular, \citet{ZBM2002} showed that the stage of period decrease lasted from around 1770
when the star oscillated with the period $\Pi_a^\star\approx 480$ d to nearly 1950 when the period
reduced to $\Pi_b^\star\approx 380$ d.
Thus, at present R~Hya is the only Mira--type variable star with observational estimates
of the period decrease time interval $\Delta t_{ab}^\star \approx 180$~yr as well as the period
values $\Pi_a^\star$ and  $\Pi_b^\star$ at the endpoints of this interval.

R~Hya is a Mira--type pulsating variable star \citep{SKDKP2017} with absorption spectrum
of class M6e \citep{M1946,M1952,M1957} corresponding to a surface carbon to oxygen abundance
ratio $\textrm{C}/\textrm{O}<1$.
At the same time detection of absorption lines of the radioactive element technetium (Tc I)
shows that R~Hya is a thermally pulsing AGB star experiencing the third dredge--up
\citep{OSh1984, L1987, LH2003}.
According to the evolutionary calculations of \citet{WZ1981} the period decrease in R~Hya
is due to changes of the stellar radius and luminosity accompanying the helium--shell flash.
Unfortunately, no detailed stellar pulsation calculations for R~Hya models have been done so far
so that the mass of R~Hya still remains unclear.

The period of radial oscillations and the stellar radius relate as $\Pi\propto R^{3/2}$ therefore
some qualitative considerations about period change can be obtained from temporal variations
of the stellar radius.
A helium--shell flash is accompanied by two consecutive phases of declining stellar radius and
luminosity \citep{WZ1981, I1982, BS1988}.
The onset of the first stage nearly coincides with the maximum peak of the helium--shell
luminosity $L_{3\alpha}$ when the hydrogen--burning shell rapidly extinguishes \citep{F2022}.
The stellar contraction ceases as soon as the radiation--diffusion wave from the helium--shell flash
propagates up to the outer convection zone.
The duration of the first radius decline $\Delta t_1$ is several dozen times shorter than that of
the second radius decline $\Delta t_2$ and both these time scales decrease with increasing
mass of the degenerate carbon core.
In particular, for AGB stars with initial masses $1.5M_\odot\le M_\mathrm{ZAMS}\le 5M_\odot$
these time intervals are in the range $10~\textrm{yr}\loa\Delta t_1\loa 300~\textrm{yr}$ and
$150~\textrm{yr}\loa\Delta t_2\loa 1.5\times 10^4$, respectively.

A first goal of the present work is to determine the evolutionary sequence with radius
decline responsible for the pulsation period decrease observed in R~Hya.
In particular, we look for a model with the pulsation period at the onset of radius decline
$\Pi\approx 480$ d and the duration of following period decrease $\approx 180$ yr.
The second goal of this work is that the selected evolutionary sequences are used for
construction of hydrodynamic models describing the evolution of the pulsation period
so that results of hydrodynamic computations can be compared with observations of the period
in R~Hya obtained in the last two centuries.
To this end we compute the non--linear stellar pulsations using the time--dependent
inner boundary conditions determined from the evolutionary calculations.
This method has been earlier employed for modelling the pulsation period decrease in the
Mira--type variable T~UMi \citep{F2022}.

\section{Stellar evolution calculations}

The present study is based on calculations of the stellar evolution from the main sequence up to
the AGB tip.
Selected models of evolutionary sequences were used to determine the initial and time--dependent
inner boundary conditions for the equations of radiation hydrodynamics describing radial
stellar oscillations.
We considered several evolutionary sequences with masses on the main sequence
$1.5M_{\sun}\le M_\mathrm{ZAMS}\le 5.0M_{\sun}$ and with the initial helium abundance $Y=0.28$.
The initial abundance of elements heavier than helium was assumed to be the same
as the solar metallicity $Z=0.014$ \citep{AGSS2009}.

Evolutionary computations were performed using the MESA program version r15140 \citep{Paxton2019}.
Convective mixing was treated following \citet{BV1958} with the mixing length to pressure
scale height ratio $\alpha_\mathrm{MLT}=1.8$.
Extra mixing at boundaries of convection zones was calculated according to \citet{H2000}
with overshooting parameter values recommended by \citet{Pignatari2016}.
The equations of nucleosynthesis were solved for the network consisting of 26 isotopes
from hydrogen ${}^1\textrm{H}$ to magnesium ${}^{24}\textrm{Mg}$.
The rates of 81 nuclear reactions were calculated with the data base REACLIB \citep{Cyburt2010}.
The mass loss rate at evolutionary stages prior to AGB was evaluated following to \citet{R1975}
with the parameter $\eta_\mathrm{R}=0.5$, whereas in calculations of the AGB stage the mass loss rate
was computed according to \citet{B1995} with the parameter $\eta_\mathrm{B}=0.05$.

\section{Evolutionary sequences for R Hya}

\begin{figure}
	\includegraphics[width=\columnwidth]{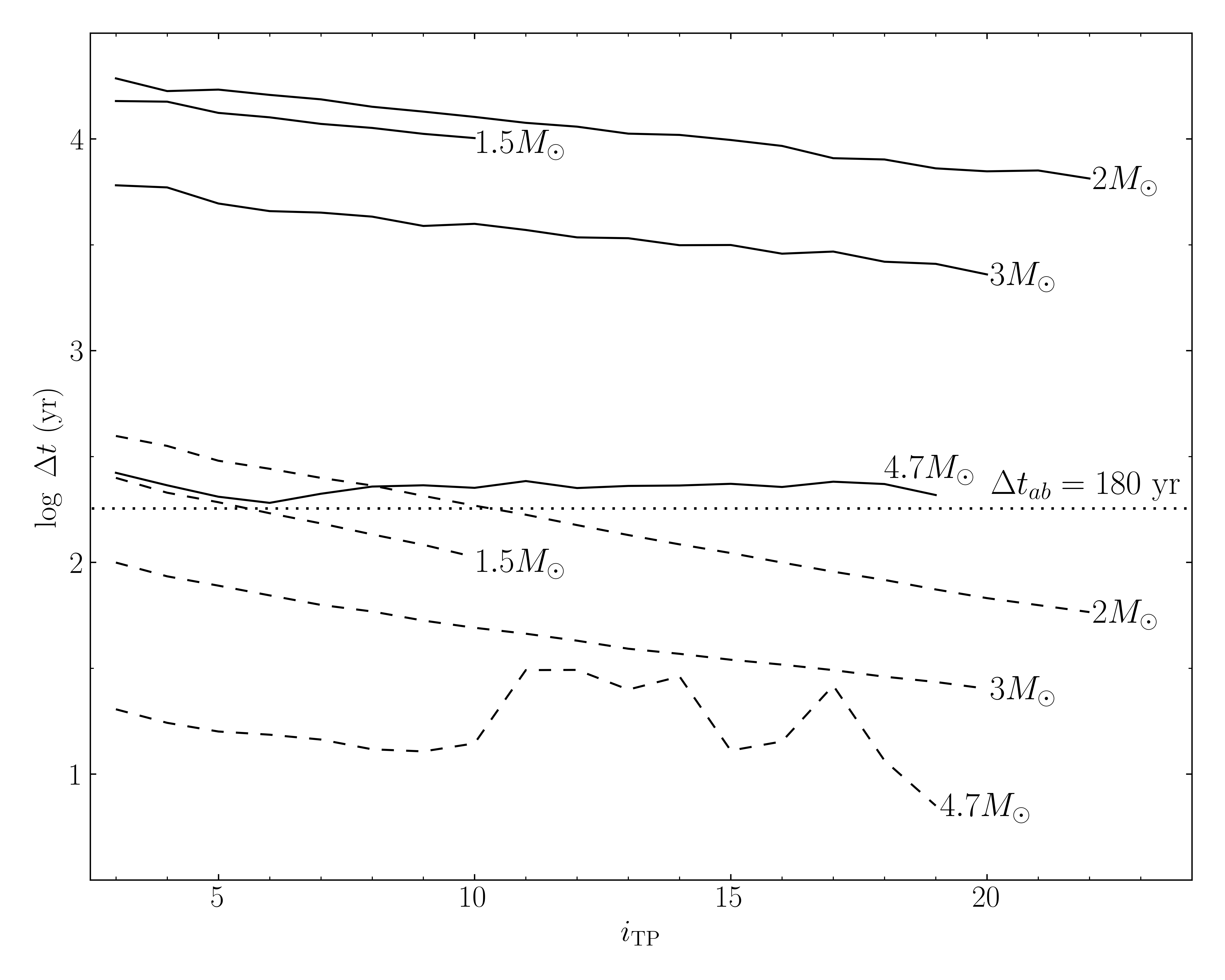}
    \caption{Duration of the first (dashed lines) and second (solid lines) radius decline
    as a function of the number of thermal pulse $i_\mathrm{TP}$.
    The initial mass $M_\mathrm{ZAMS}$ is indicated at the curves.
    The dotted line represent the time interval $\Delta t_{ab} = 180$ yr.}
    \label{fig:fig1}
\end{figure}

Fig.~\ref{fig:fig1} shows the duration of time intervals as a function of the number of the thermal
pulse $i_\mathrm{TP}$ for evolutionary sequences $1.5M_\odot\le M_\mathrm{ZAMS}\le 4.7M_\odot$.
The plots corresponding to the first and second radius decline are shown by dashed and solid
lines, respectively.
As can be seen from these plots, to explain period decrease in R~Hya we should consider
the models of evolutionary sequences $1.5M_\odot\le M_\mathrm{ZAMS}\le 2M_\odot$
for thermal pulses $4\loa i_\mathrm{TP}\loa 11$ (first radius decline) and
$M_\mathrm{ZAMS}\ge 4.7M_\odot$ for $i_\mathrm{TP}\goa 5$ (second radius decline).

\begin{figure}
	\includegraphics[width=\columnwidth]{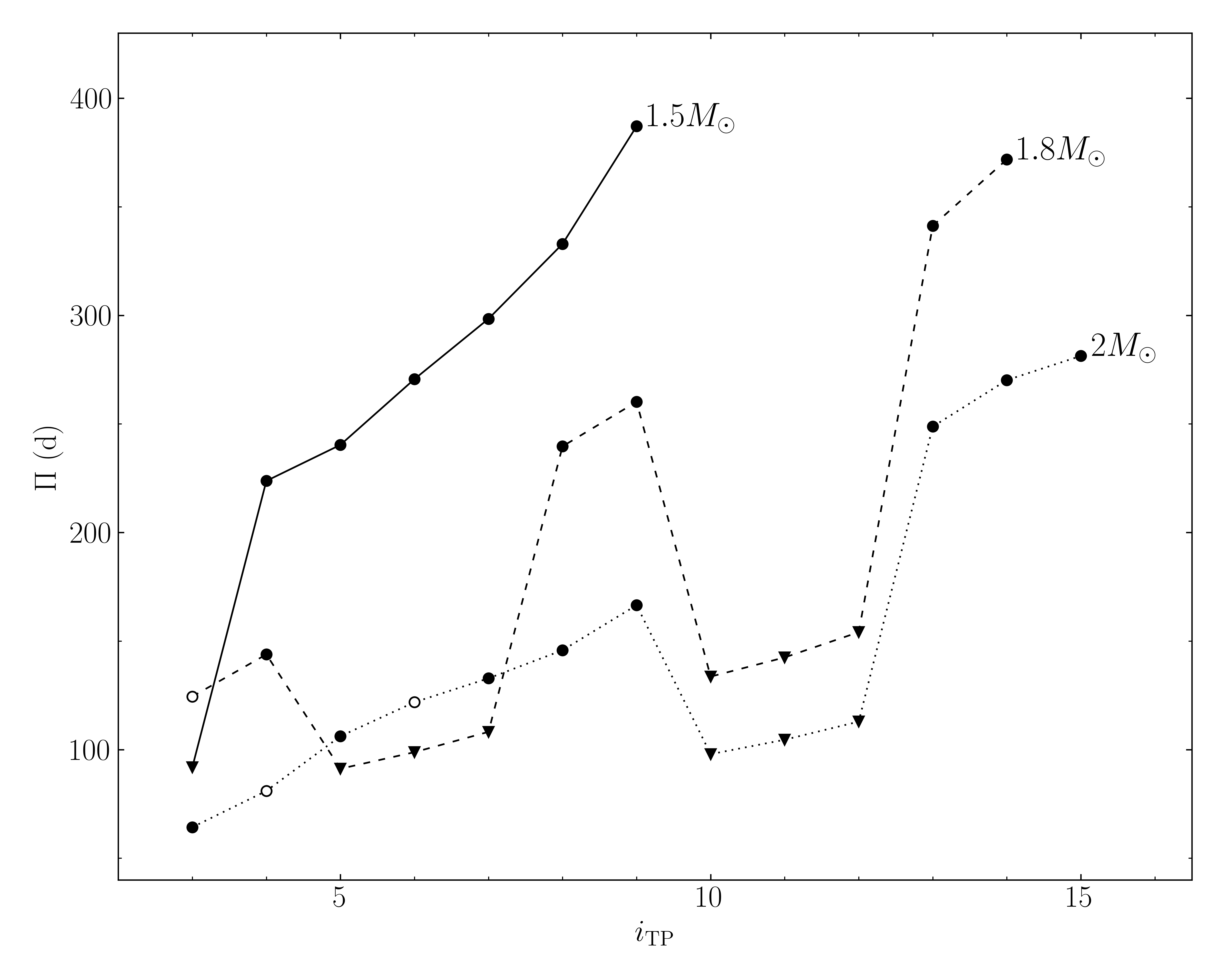}
    \caption{The pulsation period at the onset of radius decline as a function of the number
    of thermal pulse $i_\mathrm{TP}$ for evolutionary sequences $M_\mathrm{ZAMS}=1.5M_\odot$,
    $1.8M_\odot$ and $2M_\odot$.
    Filled circles and triangles correspond to the pulsation instability in the fundamental
    mode and first overtone.
    Open circles show fundamental mode periods of pulsationally stable models.}
    \label{fig:fig2}
\end{figure}

To determine the applicability of evolutionary sequences $1.5M_\odot\le M_\mathrm{ZAMS} \le 2M_\odot$
for explanation of period decrease in R~Hya we calculated the pulsation periods at
maxima of the helium--burning shell luminosity corresponding to the onset of first radius decline.
To this end we computed the self--excited radial oscilations (see below section~\ref{sec:initcond})
and evaluated the period after attainment of limiting amplitude.
Results of pulsation computations for evolutionary sequences $M_\mathrm{ZAMS}=1.5M_{\sun}$,
$1.8M_\odot$ and $2M_\odot$ are shown in Fig.~\ref{fig:fig2}, where the filled circles and
triangles represent the pulsation periods corresponding to the fundamental mode and the first
overtone.
Fundamental mode periods of three hydrodynamic models that were found to be stable against
radial oscillations are shown in Fig.~\ref{fig:fig2} by open circles.

As seen in Fig.~\ref{fig:fig2}, the fundamental mode periods of evolutionary sequences
$1.5M_\odot\le M_\mathrm{ZAMS} \le 2M_\odot$ are significantly smaller than the period
$\Pi_a\approx 480$~d corresponding to the onset of period decrease in R~Hya.
Therefore, any stellar models with period decrease during the first radius decline
can be rejected from further consideration.

\begin{figure}
	\includegraphics[width=\columnwidth]{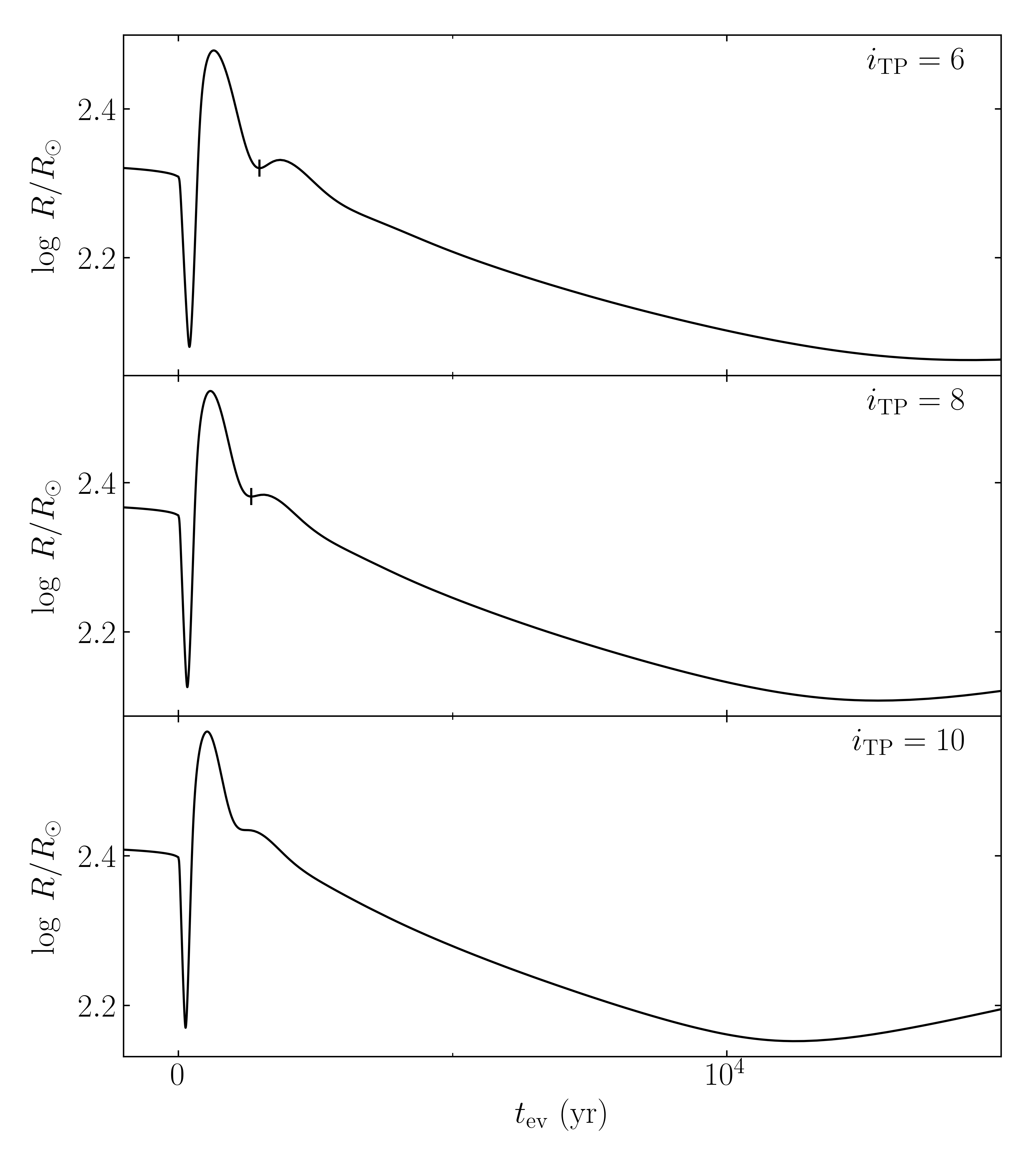}
    \caption{Time dependences of the stellar radius for the models of the evolutionary
    sequence $M=1.8M_\odot$ during the thermal pulses $i_\mathrm{TP}=6$, 8 and 10.
    Vertical ticks on the plots mark the local radius minimum prior to the secondary
    maximum of the radius.}
    \label{fig:fig3}
\end{figure}

Evolutionary calculations by \citet{VW1993} show that the luminosity and radius time dependencies
of some thermally pulsing AGB star models exhibit the bump during the second radius decline.
The bump (or the secondary maximum) is due to the thermal wave of the helium--shell flash that
reflects from the star's center.
This feature is illustrated in Fig.~\ref{fig:fig3}, where the temporal dependences of the
stellar radius in the models of the evolutionary sequence $M_\mathrm{ZAMS}=1.8M_\odot$
are shown for thermal pulses $i_\mathrm{TP}=6$, 8 and 10.
For the sake of convenience, the time $t_\mathrm{ev}$ in these plots is set to zero at
the maximum peak luminosity $L_{3\alpha}$ of the helium--shell flash.

As seen in Fig.~\ref{fig:fig3}, the time interval between the maximum radius and the local radius
minimum prior to the bump is much shorter than that of the second radius decline.
Therefore, one could attempt to construct the model with temporary cessation of radius decline
in order to reconcile observations of R~Hya with theoretical estimates of the pulsation period
at the maximum radius and the period decrease duration.

\begin{figure}
	\includegraphics[width=\columnwidth]{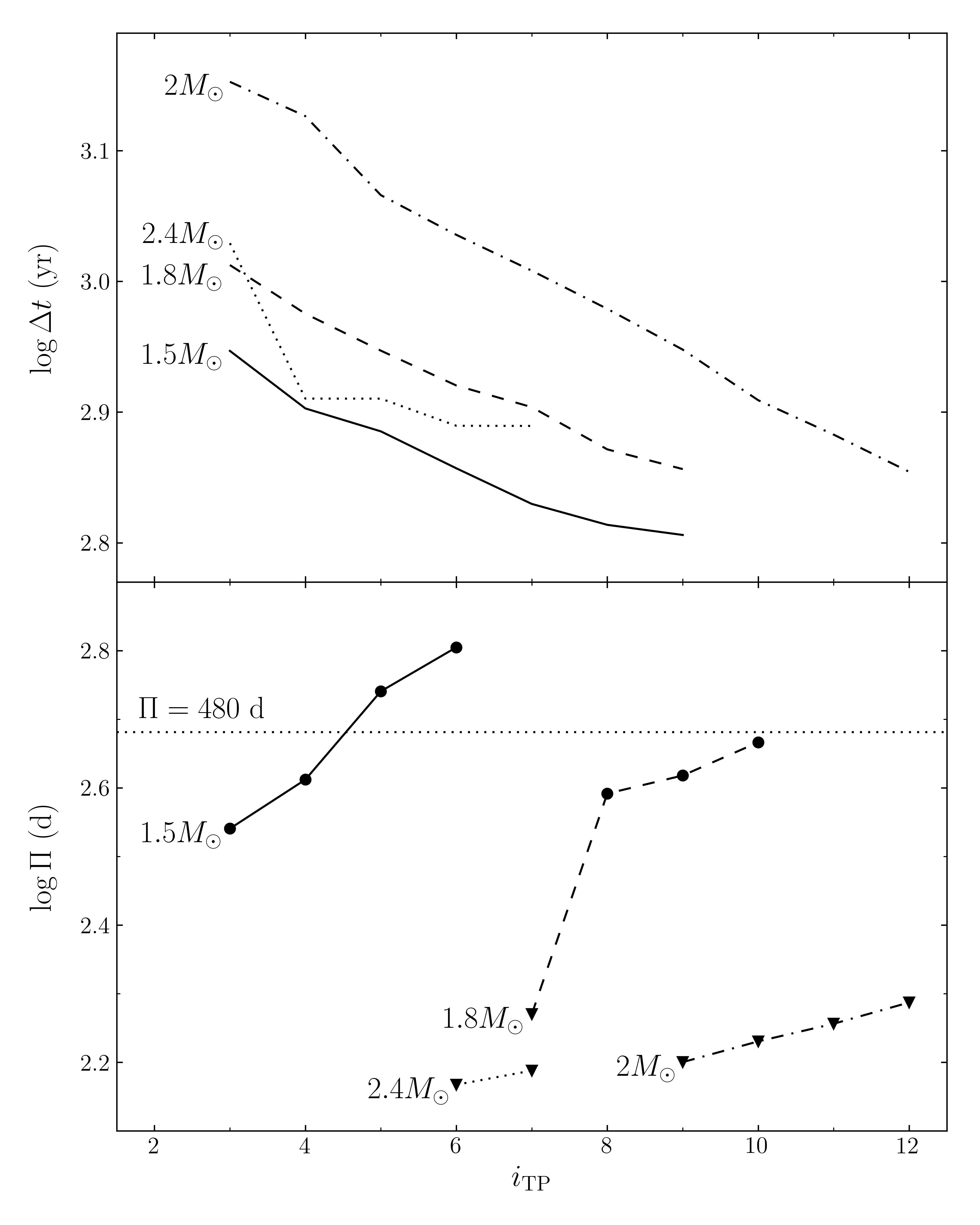}
    \caption{The time interval between the radius maximum and the local minimum due to the bump
    (upper panel) and the pulsation period at the maximum radius (lower panel) as a function
    of the number of thermal pulse $i_\mathrm{TP}$. Filled circles and triangles
    represent the fundamental mode and first overtone pulsators, respectively.}
    \label{fig:fig4}
\end{figure}

In the present study the bump during the second radius decline was found in the evolutionary
sequences with initial masses $1.5M_{\sun}\le M_\mathrm{ZAMS}\le 2.4M_{\sun}$.
Both the duration of the radius decline and the time interval
between the maximum radius and the local radius minimum prior to the bump $\Delta t$ decrease
as the star evolves and the carbon core mass grows.
Plots of the length of the time interval $\Delta t$ as a function of the number of thermal pulse
$i_\mathrm{TP}$ are shown in the upper panel of Fig.~\ref{fig:fig4}.
As seen in Fig.~\ref{fig:fig3}, the amplitude of the bump reduces with increasing $i_\mathrm{TP}$
so that finally the secondary maximum transforms into the bump with monotonic
decline of the radius where only the second time derivative of the radius changes the sign.
The end--points of the plots in the upper panel of Fig.~\ref{fig:fig4} correspond to the
last secondary maximum before its disappearance during the following thermal pulse.
The lower limit of the time interval ranges from $\Delta t=640~\textrm{yr}$ for
$M_\mathrm{ZAMS}=1.5M_{\sun}$ to $\Delta t=780~\textrm{yr}$ for $M_\mathrm{ZAMS}=2.4M_{\sun}$.

Envelopes of the stellar models at the maximum radius after the helium flash are under thermal
equilibrium and their pulsation periods we evaluated using the computational procedure
described above.
Because in each evolutionary sequence the time interval $\Delta t$ decreases monotonically with
increasing $i_\mathrm{TP}$, of most interest are the pulsation models in the vicinity of the
end--point of the plots in the upper panel of Fig.~\ref{fig:fig4}.

Results of pulsation calculations are presented in the lower panel of Fig.~\ref{fig:fig4}.
In the evolutionary sequence $M_\mathrm{ZAMS}=1.5M_{\sun}$ the model corresponding to $i_\mathrm{TP}=6$
pulsates in the fundamental model with period $\Pi=550$ d whereas the time interval between
the radius maximum and the local minimum due to the bump is $\Delta t = 770$ yr.
On the other hand, the models of evolutionary sequences $M_\mathrm{ZAMS}=2M_{\sun}$
and  $M_\mathrm{ZAMS}=2.4M_{\sun}$ pulsate in the first overtone with periods shorter than 200~d.

Thus, an assumption that the period decrease in R~Hya occured during the second radius decline
from the maximum radius to the local radius minimum prior to the secondary maximum
is not compatible with the twin requirements that the maximum
pulsation is about 480 d and the duration of period decline lasts
only about 180 yr.
To confirm this conclusion we can consider the model of the evolutionary sequence
$M_\mathrm{ZAMS}=1.8M_{\sun}$ with period at the maximum stellar radius $\Pi_a=465$ d
($i_\mathrm{TP}=10$) which has less deviation from the period value $\Pi_a^\star=480$ d.
As seen in the lower panel of Fig.~\ref{fig:fig3}, the secondary maximum in the plot for this
model disappears but location of the bump in the plot nearly corresponds to the time interval
$\Delta t\approx 700$~yr.

\begin{figure}
	\includegraphics[width=\columnwidth]{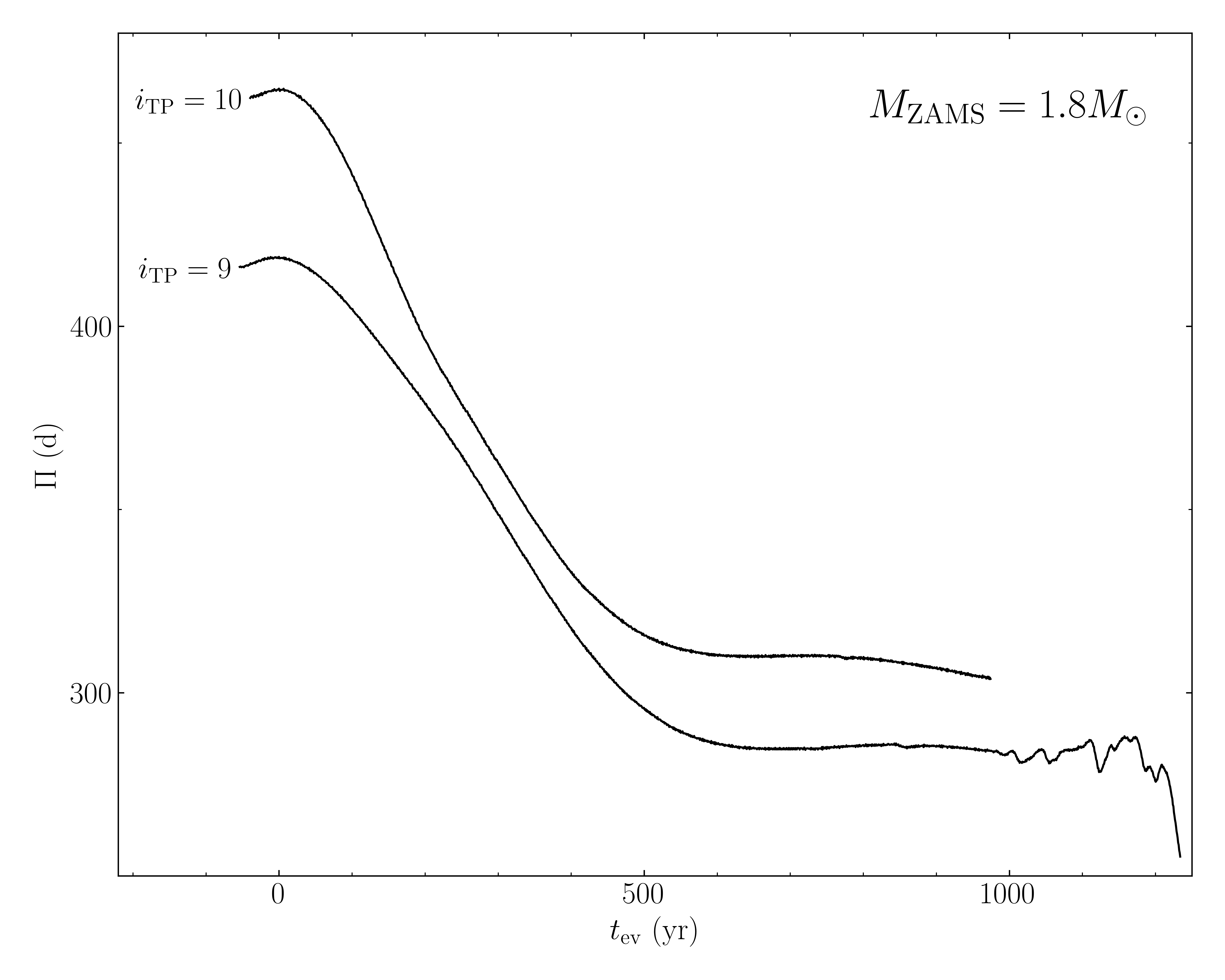}
    \caption{Evolution of the pulsation period in the hydrodynamic models of the evolutionary
    sequence $M_\mathrm{ZAMS}=1.8M_{\sun}$ at thermal pulses $i_\mathrm{TP}=9$ and $i_\mathrm{TP}=10$.}
    \label{fig:fig5}
\end{figure}

Fig.~\ref{fig:fig5} presents time variations of the pulsation period $\Pi(t_\mathrm{ev})$ during
radius decline for the thermal pulses $i_\mathrm{TP}=9$ and $i_\mathrm{TP}=10$ in the evolutionary
sequence $M_\mathrm{ZAMS}=1.8M_{\sun}$.
To obtain these dependencies we solved the initial--value problem for equations of hydrodynamics
with the inner boundary conditions determined from evolutionary models.
The method of calculations is described in \citet{F2022}.

As seen from plots in Fig.~\ref{fig:fig5}, the duration of the period decrease is
$\Delta t\approx 500$~yr.
The pulsation period during the temporary cessation of the radius decline
($600~\textrm{yr}\loa t_\mathrm{ev}\loa 800~\textrm{yr}$) is $\Pi_b=310$ d and
the period ratio $\Pi_a/\Pi_b = 1.50$ is significantly larger than that in R~Hya
($\Pi_a^\star/\Pi_b^\star = 1.26$).
Thus, the AGB star models with initial mass $M_\mathrm{ZAMS} < 3M_\odot$ cannot be reconciled with
observations of R~Hya and therefore we have to consider the models of more massive stars.

As was
shown in our previous paper \citep{F2023}, the duration of the second radius decline
$\Delta t_{ab}$
deviates less from the duration of
period decrease in R~Hya for models of the evolutionary
sequence with initial mass $M_\mathrm{ZAMS} \approx 4.8_{\sun}$.
Results of more detailed calculations carried out in the present study are summarized in
Table~\ref{tab:table1}, where the values of $\Delta t_{ab}$ corresponding to the thermal
pulses $4\le i_\mathrm{TP} \le 9$ are given for models of evolutionary sequences
$4.5M_{\sun} \le M_\mathrm{MZAMS} \le 5.0M_{\sun}$.
As seen in Table~\ref{tab:table1}, the duration of the second contraction phase decreases with
increasing stellar mass and is insignificantly sensitive to $i_\mathrm{TP}$.
Bearing in mind that the duration of period decrease in R~Hya is
$\Delta t_{ab}^\star \approx 180$ we can conclude that the initial masses of the most appropriate
evolutionary sequences are $4.7M_{\sun}\le M_\mathrm{ZAMS}\le 4.9M_{\sun}$.

\begin{table}
	\centering
	\caption{Duration of the second contraction phase in thermal pulses $4\le i_\mathrm{TP} \le 9$
	of evolutionary sequences $4.5M_{\sun}\le M_\mathrm{ZAMS}\le 5.0M_{\sun}$.}
	\label{tab:table1}

\begin{tabular}{lcccccr}
\hline
$M_\mathrm{ZAMS}$ & \multicolumn{6}{c}{$\Delta t_{ab}\ (\textrm{yr})$} \\
$M_{\sun}$ &      4 &      5 &      6 &      7 &      8 &      9 \\
\hline
     4.5 &  295 &  257 &  234 &  226 &  260 &  268 \\
     4.6 &  243 &  228 &  209 &  218 &  248 &  257 \\
     4.7 &  231 &  204 &  191 &  211 &  228 &  231 \\
     4.8 &  219 &  197 &  173 &  189 &  188 &  187 \\
     4.9 &  206 &  171 &  151 &  182 &  182 &  185 \\
     5.0 &  195 &  160 &  153 &  157 &  155 &  157 \\
\hline
\end{tabular}
\end{table}

\begin{figure}
	\includegraphics[width=\columnwidth]{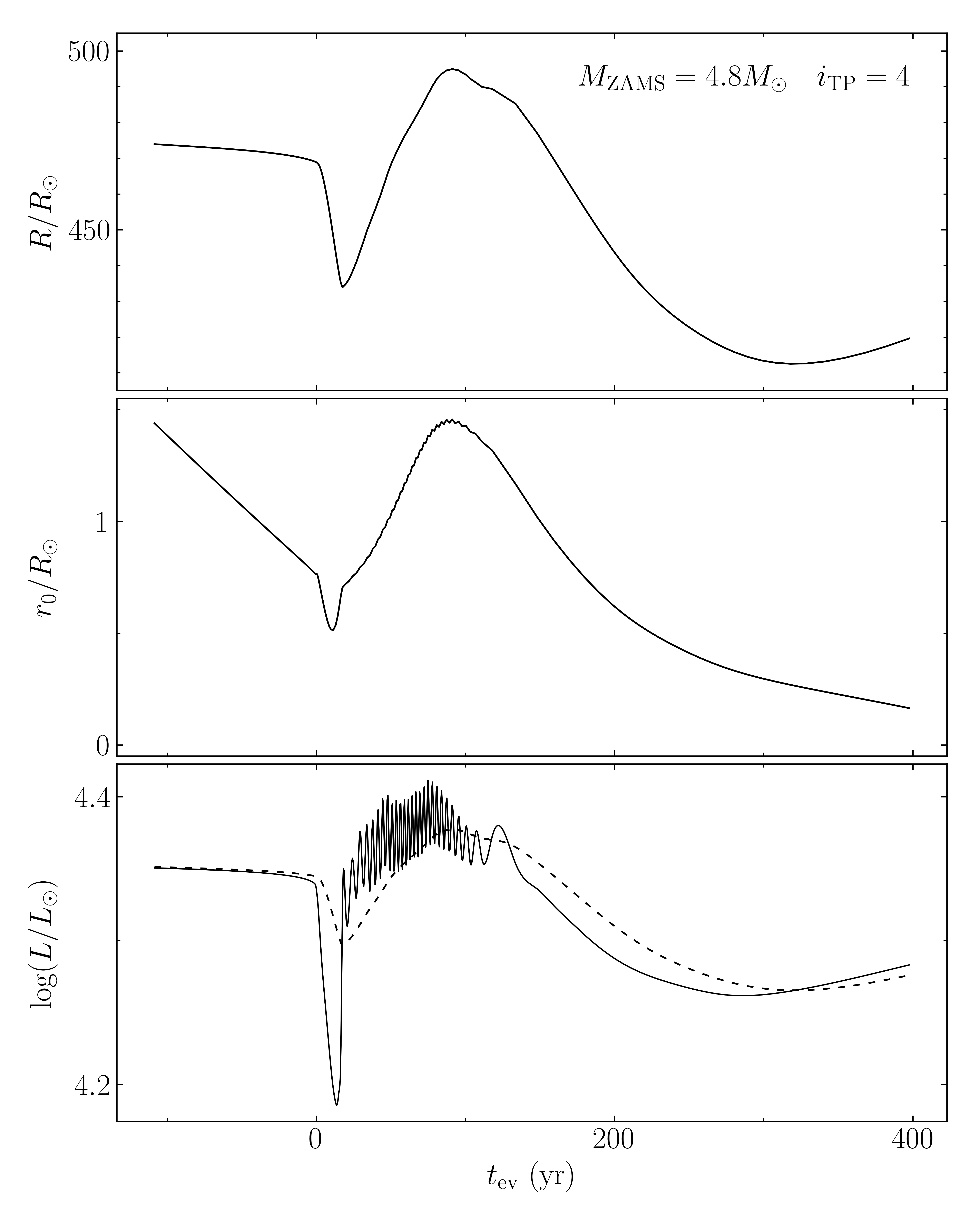}
    \caption{Variations with time of the surface radius $R$ (upper panel) and
             the radius at the inner boundary $r_0$ (middle panel)
             for the evolutionary sequence $M_\mathrm{ZAMS}=4.8M_{\sun}$
             during the thermal pulse $i_\mathrm{TP}=4$.
             In the bottom panel, the inner boundary condition $L_0$ (solid line)
             and the stellar luminosity $L$ (dashed line) as a function of the age
             $t_\mathrm{ev}$.}
    \label{fig:fig6}
\end{figure}

The time--dependent inner boundary conditions are illustrated in Fig.~\ref{fig:fig6}.
The upper panel of Fig.~\ref{fig:fig6} shows the temporal variation of the surface radius
of the evolutionary model $R$
in the time interval spanned by hydrodynamic computations for the model of the evolutionary
sequence $M_\mathrm{ZAMS}=4.8M_{\sun}$ during the 4--th thermal pulse.
For the sake of convenience the evolution time $t_\mathrm{ev}$ is set to zero at the maximum
peak luminosity $L_{3\alpha}$ of the helium--shell flash.
The inner boundary of the hydrodynamic model is set at the Lagrangean mass coordinate $M_0 = 0.187M$,
where $M=4.605M_{\sun}$ is the mass of the AGB star model.
The time--dependent inner boundary conditions $r_0(t_\mathrm{ev})$ and $L_0(t_\mathrm{ev})$ are
determined by non--linear interpolation with respect to the mass coordinate $M_r$ and are
shown in the middle and bottom panels, respectively.
The dashed line in the bottom panel represents the surface luminosity of the evolution model.

The most striking feature in Fig.~\ref{fig:fig6} is the presence of oscillations
of the inner boundary luminosity $L_0(t_\mathrm{ev})$ during expansion of the envelope
within the time interval $15~\textrm{yr}\loa t_\mathrm{ev}\loa 90~\textrm{yr}$.
These oscillations arise due to both the small-amplitude variations of the energy released
by hydrogen burning and the close location of the inner boundary to the hydrogen--shell
source.
For example, in the model shown in Fig.~\ref{fig:fig6} the mass difference between the
hydrogen--shell source and the inner boundary is $\delta M_r\sim 10^{-3}M_{\sun}$.
To diminish the amplitude of oscillations in $L_0$ we carried out the evolutionary
computations with the larger number of mass zones $N$.
The plot in the lower panel of Fig.~\ref{fig:fig6} corresponds to $N\approx 8\times 10^4$
so that oscillations of $L_0$ with amplitude $\delta L_0/L_0\loa 10\%$ do not affect
the surface luminosity.

\section{Initial conditions for hydrodynamic models}
\label{sec:initcond}

Calculations of hydrodynamic stellar models with evolving pulsation period are done using
the initial conditions in the form of the limiting amplitude stellar pulsation model
computed with the time--independent inner boundary conditions
\begin{equation}
\frac{\partial r_0}{\partial t} = \frac{\partial L_0}{\partial t} = 0 .
\end{equation}
The necessary condition for correct application of the theory of stellar pulsation is
that the evolutionary stellar model is under hydrostatic and thermal equilibrium.
All models of evolutionary sequences are under hydrostatic equilibrium  and at the same time
may be in thermal imbalance during stellar contraction or expansion.
Therefore, the selection of appropriate evolutionary models should be based on the estimate
of the deviation from thermal equilibrium.
For the spherically symmetric stellar envelope without nuclear energy sources
the deviation from thermal equilibrium can be expressed as
\begin{equation}
\delta_\mathrm{L} = \max_{1\le j\le N} |1 - L_j/L_0| ,
\end{equation}
where $L_j$ is the total (radiative plus convective) luminosity at the $j$--th mass zone,
$L_0$ is the luminosity at the inner boundary, $N$ is the number of mass zones in
the hydrodynamic model.
Evolution models with the age $t_\mathrm{ev}\approx -100$~yr, where $t_\mathrm{ev} = 0$
at the maximum peak luminosity $L_{3\alpha}$, were found to satisfy the condition
$\delta_\mathrm{L}\la 10^{-3}$ corresponding to negligible thermal imbalance in the stellar
envelope.

The Lagrangean mesh of the initial hydrodynamic model was determined for the outer layers
with the radius at the innermost mass zone $r_0\sim 10^{-3}R$, where $R$ is the radius
of the outer boundary of the evolution model.
All hydrodynamic models were computed with $N=900$ mass zones, where the mass intervals of
800 outer zones increase geometrically inwards whereas the mass intervals of 100 inner zones
decrease.
This approach allowed us to obtain a better approximation in the inner layers of the stellar
envelope where the pressure and temperature gradients sharply change.
The variables specified on the Lagrangean mesh of the initial hydrodynamic model were
evaluated using the non--linear interpolation of stellar evolution model data with respect
to the mass coordinate $M_r$.

To calculate the self--excited non--linear stellar oscillations we solved the equations
of radiation hydrodynamics and time--dependent convection where the transport equations for
convective mixing were used in the form by \citet{K1986}.
The full system of the equations of hydrodynamics and the choice of parameters are discussed
in our preceding paper \citep{F2015}.

\begin{figure}
	\includegraphics[width=\columnwidth]{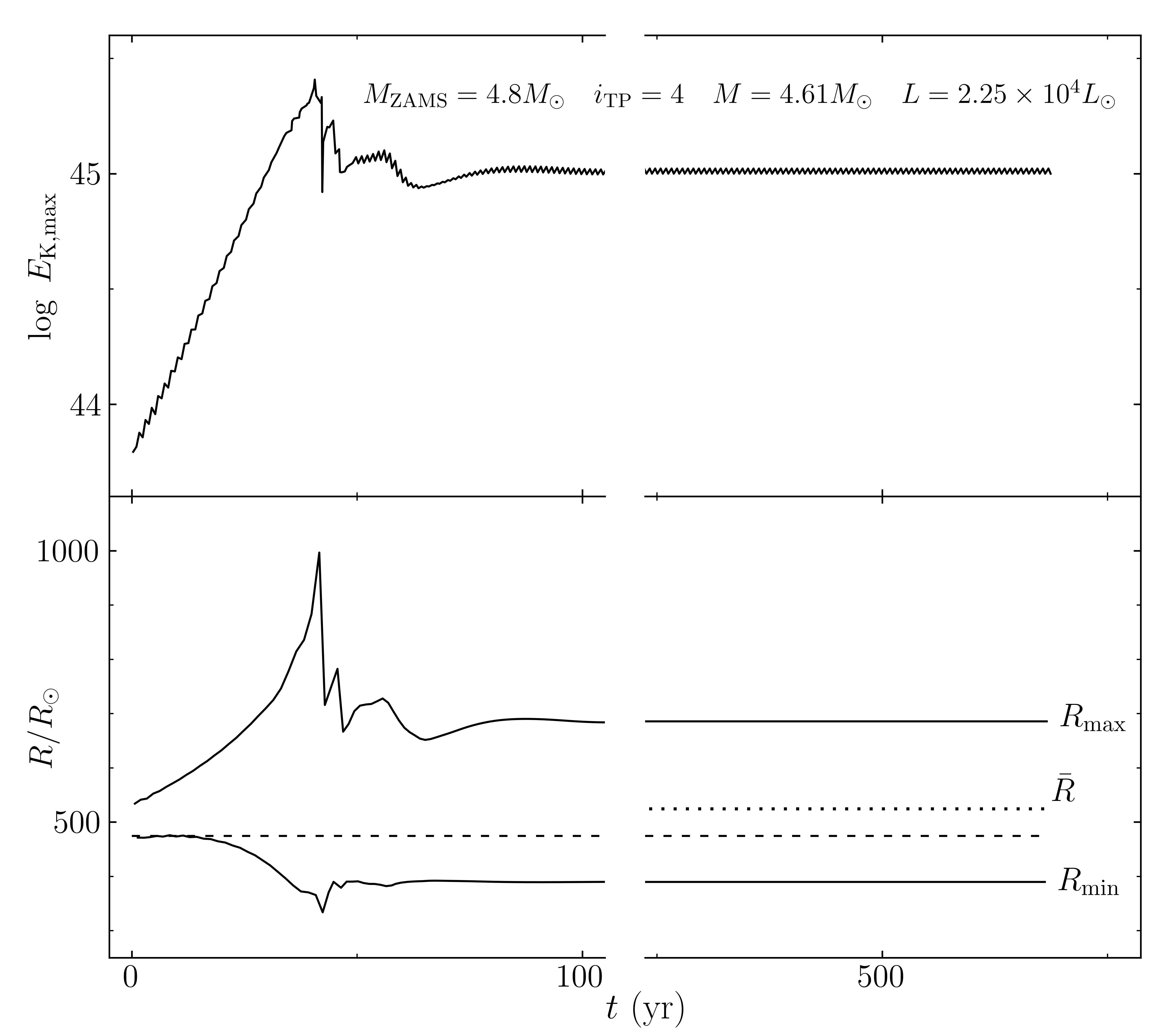}
    \caption{The maximum kinetic energy $E_\mathrm{K,\max}$ versus time $t$ (top panel) for the
             hydrodynamic model of the evolutionary sequence $M_\mathrm{ZAMS}=4.8M_{\sun}$
             prior to the thermal pulse $i_\mathrm{TP}=4$.
             In the bottom panel, the maximum and minimum radii $R_{\max}$ and $R_{\min}$
             of the outer boundary of the hydrodynamic model (solid lines).
             The dotted and dashed lines represent the mean radius over the pulsation period
             $\bar{R}$ and the radius of the evolution model, respectively.}
    \label{fig:fig7}
\end{figure}

\begin{figure}
	\includegraphics[width=\columnwidth]{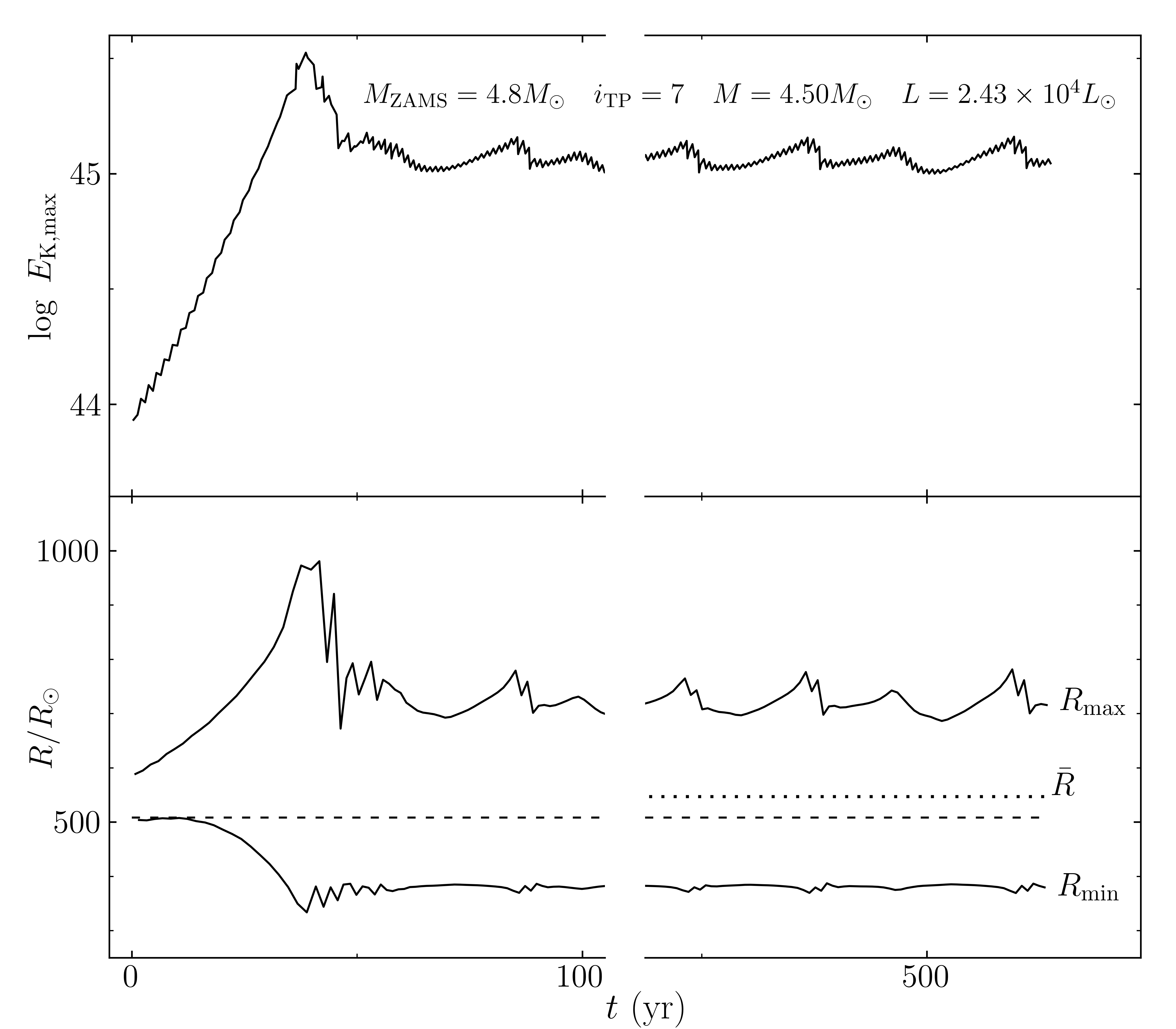}
    \caption{Same as Fig.~\ref{fig:fig7}, but for the hydrodynamic model prior to the
             thermal pulse $i_\mathrm{TP}=7$.}
    \label{fig:fig8}
\end{figure}

Results of calculations of the self--excited radial oscillations
with static inner boundary conditions
are illustrated in
Figs.~\ref{fig:fig7} and \ref{fig:fig8} for two models of the evolutionary sequence
$M_\mathrm{ZAMS}=4.8M_{\sun}$ prior to the thermal pulses $i_\mathrm{TP}=4$ and
$i_\mathrm{TP}=7$, respectively.
The plots in the left and right
sections of the figures
correspond to the stages prior to and after
attainment of limiting amplitude.
In upper panels we give the plots of the maximum values of the kinetic energy $E_\mathrm{K,\max}$
which is reached twice per pulsation period.
Both models have the close values of periods and growth rates:
$\Pi=441$~d, $\eta = \Pi d\ln E_\mathrm{K,\max}/d t = 0.11$  for $i_{TP}=4$ and
$\Pi\approx 459$~d, $\eta = 0.12$ for $i_{TP}=7$
but at the same time they demonstrate different behaviour after the limiting amplitude attainment.
This due to the fact that R~Hya models are at the edge between the regular and
numerically over--driven
stellar pulsations.

Attainment of the limiting amplitude occurs not only due to saturation of the $\kappa$--mechanism
in the hydrogen ionization zone but also because of kinetic energy dissipation by shock waves.
The latter mechanism becomes significant at the large--amplitude oscillations.
Moreover, periodic shock waves accompanying non--linear stellar pulsations are responsible for
distension of the stellar envelope\citep{W2000}.
As seen in Figs.~\ref{fig:fig4} and \ref{fig:fig5}, the radial amplitude at the outer boundary
is $\Delta R/R \approx 0.5$ whereas the ratio of the mean radius of the pulsating star to
the radius of the evolution model is $\bar{R}/R_\mathrm{ev}\approx 1.1$.
The lack of any secular change in the mean radius $\bar{R}$ indicates the absence of entropy changes
arising due to the thermal imbalance in the initial stellar model \citep{YaT1996}.

\section{Evolution of the pulsation period}

Solution of the radiation hydrodynamics equations based on the time--dependent inner boundary
conditions is almost consistent with results of stellar evolution computations.
The only exception is that the non--linear stellar pulsations are calculated for the
constant mass of the hydrodynamic model, whereas the mass of the evolution model gradually
decreases.
However hydrodynamic computations are carried out for time intervals as long as 500~yr
so that the mass difference between the evolutionary and hydrodynamic models is negligible
and does not exceed 0.1\%.

\begin{figure}
	\includegraphics[width=\columnwidth]{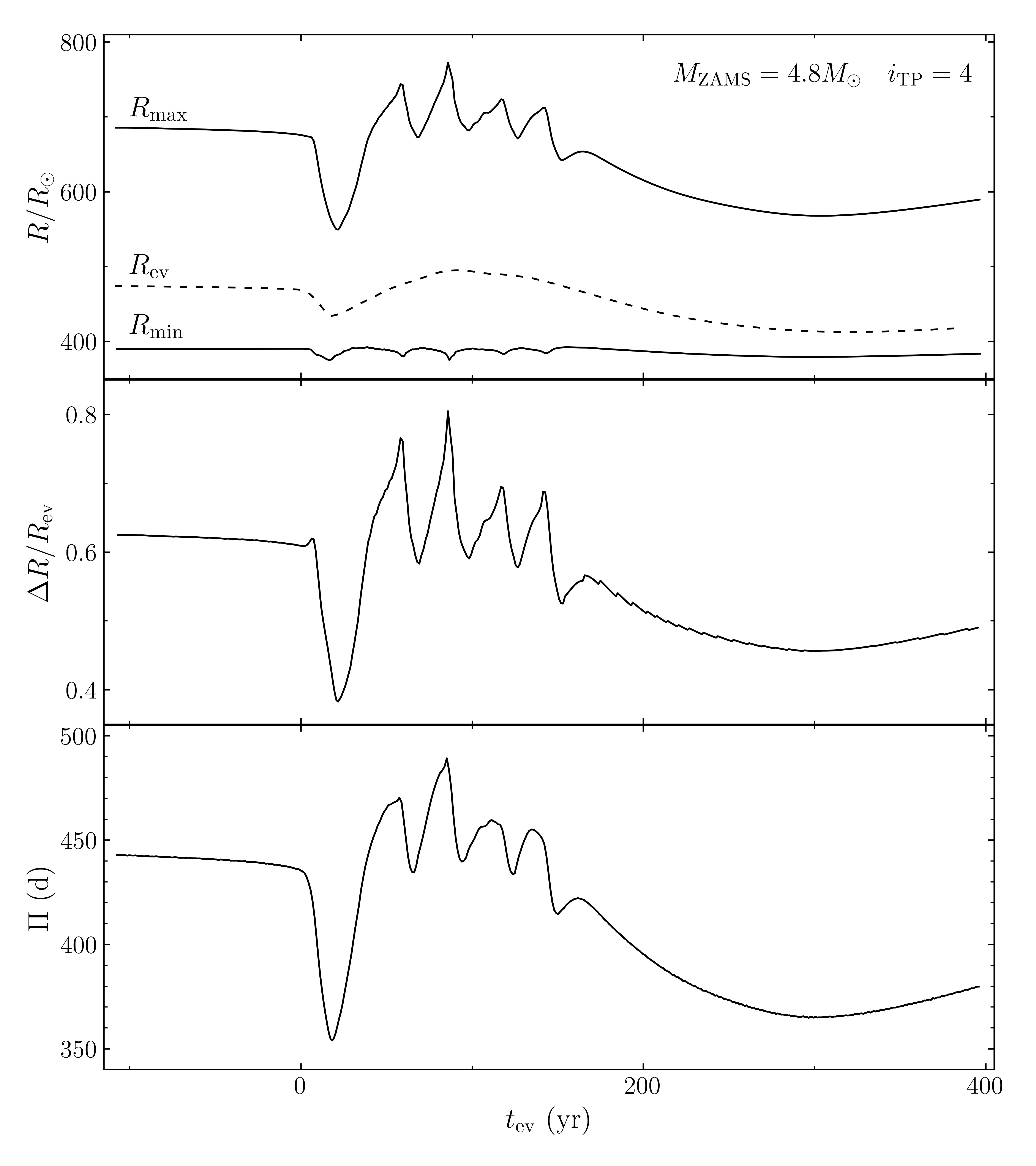}
    \caption{Upper panel: variation with evolution time $t_\mathrm{ev}$ of maximum and minimum radii
            of the hydrodynamic model with initial mass $M_\mathrm{ZAMS}=4.8M_{\sun}$ during
            the thermal pulse $i_\mathrm{TP=4}$ (solid lines).
            The dashed line shows the time variation of the surface radius of the evolution model.
            Middle and bottom panels: the pulsation amplitude and pulsation period versus evolution time.}
    \label{fig:fig9}
\end{figure}

The top panel of Fig.~\ref{fig:fig9} shows the maximum and the minimum radii $R_{\max}$ and $R_{\min}$
as a function of evolution time $t_\mathrm{ev}$ for the hydrodynamic model
corresponding to the evolutionary sequence
$M_\mathrm{ZAMS}=4.8M_{\sun}$ during the thermal pulse $i_\mathrm{TP}=4$.
The dashed line in the upper panel represents the temporal variation of the radius $R_\mathrm{ev}$
of the evolution model.

The middle panel of Fig.~\ref{fig:fig9} shows the pulsation amplitude at the outer boundary
of the hydrodynamic model $\Delta R$ expressed in units of $R_\mathrm{ev}$.
As can be seen, the radial pulsations become
numerically over--driven
near the maximum of $R_\mathrm{ev}$
within the time interval $30~\textrm{yr}\loa t_\mathrm{ev}\loa ~170~\textrm{yr}$
when the radiation diffusion wave generated by the helium--shell flash reaches the convection envelope
of the pulsating star and the pulsation amplitude becomes as large as $\Delta R > 0.7 R_\mathrm{ev}$.

Semi--regular variations of $\Pi$ with the amplitude $\loa 10\%$ and the cycle length
ranging from 20 to 30~yr (the bottom panel of Fig.~\ref{fig:fig9}) are due to the large--amplitude
changes of the stellar radius because the pulsation period $\Pi$ is nearly proportional to $R^{3/2}$.
It should be noted that all hydrodynamic models computed in the present study exhibit the similar
behaviour of the pulsation period.

\begin{figure}
	\includegraphics[width=\columnwidth]{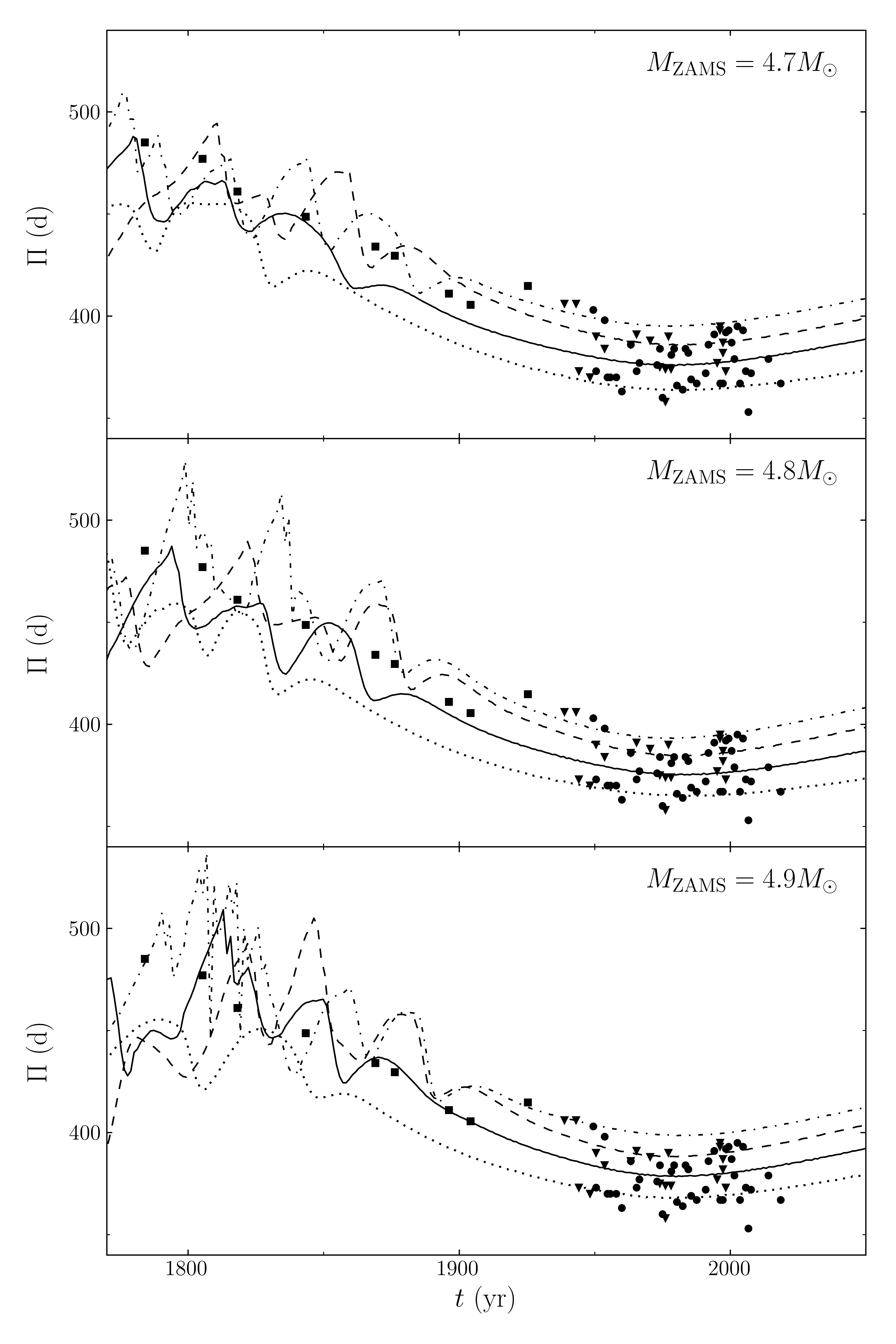}
    \caption{Pulsation period as a function of time for evolutionary sequences $M_\mathrm{ZAMS}=4.7M_{\sun}$,
    $4.8M_{\sun}$ and $4.9M_{\sun}$ for thermal pulses $i_\mathrm{TP}=4$ (dotted lines),
    $i_\mathrm{TP}=5$ (solid lines), $i_\mathrm{TP}=6$ (dashed lines) and $i_\mathrm{TP}=7$ (dashed--dotted lines).
    Observational estimates of the period for R~Hya are plotted with filled squares \citep{P1936},
    filled circles (using light curve data taken from AAVSO) and filled triangles (from the card catalogue of
    variable stars of the Sternberg Astronomical Institute).}
    \label{fig:fig10}
\end{figure}

Of most interest for comparison with observations of R~Hya is the time interval spanning over
$\approx 200$~yr between the maximum and minimum values of the period.
Fig.~\ref{fig:fig10} shows the temporal dependencies of the pulsation period for hydrodynamic models
of the evolutionary sequences $M_\mathrm{ZAMS}=4.7M_{\sun}$, $4.8_{\sun}$ and $4.9M_{\sun}$
during the thermal pulses $4\le i_\mathrm{TP}\le 7$.
In each panel of this figure we plot observational estimates of the period from \citet{P1936},
from the light curve data base of the AAVSO and from the card catalogue of variable stars of
the Sternberg Astronomical Institute.

It should be noted that in order to compare the results of computations  with observations
we arbitrarily shifted the plots of $\Pi(t)$ along the horizontal axis and as can be seen
in Fig.~\ref{fig:fig10}, the best agreement with observations is obtained for the minimum
of $\Pi(t)$ in $\approx 1980$.
At the same time one should bear in mind that the localization of the period minimum is quite
uncertain because of very slow period change near the minimum.
For example, decrease of the period dependence $\Pi(t)$ before its minimum be nearly 5\%
takes place during nearly 70 yr.

Hydrodynamic models of evolutionary sequences represented in Fig.~\ref{fig:fig10} are not very sensitive
to the initial stellar mass $M_\mathrm{ZAMS}$ but at the same time show a notable dependence on
the number of the thermal pulse $i_\mathrm{TP}$ because of the gradually increasing radius and luminosity
as a star evolves along AGB.
As can be seen in \ref{fig:fig10}, the best agreement with observations is obtained for
$5\le i_\mathrm{TP}\le 6$.

Parameters of hydrodynamic models that agree with observations age listed in
Table~\ref{tab:table2}, where in the third column we give the average period change rate $\dot\Pi$
evaluated by the linear fit of $\Pi(t)$ within the time interval $1780 \le t \le 1950$.
During the same time interval the period of R~Hya declined
almost linearly with rate $\approx -0.58~\textrm{d\:yr}^{-1}$ \citep{ZBM2002}.

\begin{table}
	\centering
	\caption{Parameters of hydrodynamic models at the minimum pulsation period $\Pi_b$.}
	\label{tab:table2}
\tabcolsep=4pt
\begin{tabular}{lcccccccr}
\hline
$M_\mathrm{ZAMS}$ & $i_\mathrm{TP}$ & $\dot\Pi$ & $M$ & $L$ & $R_\mathrm{ev}$ & $\bar{R}$ & $\Pi_b$ & C/O \\
$M_{\sun}$ & & $\text{d~yr}^{-1}$ & $M_{\sun}$ & $10^4 L_{\sun}$ & $R_{\sun}$ & $R_{\sun}$ & d & \\
\hline
4.7 & 5 & -0.57 &   4.50 & 1.84 &   410 &   469 &   364 & 0.38 \\
4.7 & 6 & -0.53 &   4.43 & 1.88 &   427 &   486 &   386 & 0.42 \\
4.8 & 5 & -0.70 &   4.58 & 1.88 &   422 &   482 &   375 & 0.37 \\
4.8 & 6 & -0.42 &   4.54 & 1.92 &   430 &   489 &   385 & 0.40 \\
4.9 & 5 & -0.57 &   4.67 & 1.95 &   428 &   487 &   378 & 0.37 \\
4.9 & 6 & -0.36 &   4.64 & 1.99 &   436 &   495 &   388 & 0.40 \\
\hline
\end{tabular}
\end{table}

The mass $M$, luminosity $L$ and radius $R_\mathrm{ev}$ of the evolution model given in
Table~\ref{tab:table2} correspond to the phase of the minimum pulsation period $\Pi_b$.
The significant difference between the average radius of the hydrodynamic model $\bar{R}$ and
the radius of the evolution model $R_\mathrm{ev}$ is due to distension of the stellar envelope
because of periodic shock waves accompanying the large--amplitude oscillations.
In the last column of Table~\ref{tab:table2} we give the surface carbon--to--oxygen number ratio C/O.
Just before the first helium--shell flash this ratio is C/O=0.33 for all evolutionary sequences
so that the computed hydrodynamic models represent the oxygen--rich Mira--type stars at the beginning
of the third dredge--up.

\section{Conclusions}

Solution of the radiation hydrodynamics equations with time--dependent inner boundary conditions allows
us to conclude that nearly 200 years ago the amplitude of stellar oscillations in R~Hya reached the
maximum value
due to substantial increase in the both surface radius and luminosity.
The large--amplitude stellar oscillations give rise to the shock wave driven mass loss and
it is of interest to note that during decline of the stellar radius, luminosity and pulsation amplitude
between 1770 and 1950 the mass loss rate decreased by a factor of $\approx 20$ \citep{DBRRHK2008}.
Moreover, the large--amplitude pulsations taking place two centuries ago seem to be responsible
for the formation of the cold dust shell surrounding R~Hya \citep{HIKB1998}.

As seen in Fig.~\ref{fig:fig7}, a major disagreement between the calculated temporal dependencies
of the pulsation period $\Pi(t)$ and observational period estimates for R~Hya corresponds to observations
conducted earlier than
1900 when the pulsation amplitude was significantly larger than nowadays.
Unfortunately, the large--amplitude
oscillations
of the hydrodynamic models
cannot be corroborated by observations
of R~Hya because they were rare and inexact.

Observations of R~Hya are well fitted with the temporal dependencies
calculated for the hydrodynamic models of the evolutionary sequence $M_\mathrm{ZAMS}=4.7M_{\sun}$
during the thermal pulses $i_\mathrm{TP}=5$ and $i_\mathrm{TP}=6$.
Moreover, the mean radius of this hydrodynamic model near the minimum of the period
(see Table~\ref{tab:table2}) has the least deviation from the observational estimate of the radius
$R=442R_{\sun}\pm 65R_{\sun}$ obtained from interferometric angular diameter measurements
of R~Hya \citep{HST1995}.

\section*{Acknowledgements}

The author is indebted to amateur astronomers of the AAVSO for their observations of R~Hydrae.
The author also expresses his gratitude to Prof. N.N. Samus', the head of the GCVS Research Group
of the Sternberg Astronomical Institute, for his help in accessing the historical card catalogue
of variable stars.
The author is grateful to the anonymous referee for his meticulous review of this paper
and the valuable comments.

\section*{Data Availability}

The data underlying this article will be shared on reasonable request to the corresponding author.


\bibliographystyle{mnras}
\bibliography{fadeyev}


\bsp	
\label{lastpage}
\end{document}